\documentclass[english,preprint,aip,jcp]{revtex4-1}
\usepackage[T1]{fontenc}
\usepackage[latin9]{inputenc}
\usepackage{amsmath}
\usepackage{amssymb}
\usepackage{graphicx}

\makeatletter

\newcommand{\lyxdot}{.}

\usepackage{enumitem}
\usepackage{amsmath}
\usepackage{accents}
\newlength{\dhatheight}
\newcommand{\doublehat}[1]{%
    \settoheight{\dhatheight}{\ensuremath{\hat{#1}}}%
    \addtolength{\dhatheight}{-0.15ex}%
    \widehat{\vphantom{\rule{1pt}{\dhatheight}}%
    \smash{\widehat{#1}}}}

\makeatother

\usepackage{babel}
\begin{document}

\title{Ehrenfest+R Dynamics II: A Semiclassical QED Framework for Raman
Scattering}

\author{Hsing-Ta Chen}
\email{hsingc@sas.upenn.edu}

\selectlanguage{english}%

\affiliation{Department of Chemistry, University of Pennsylvania, Philadelphia,
Pennsylvania 19104, U.S.A.}

\author{Tao E. Li}

\affiliation{Department of Chemistry, University of Pennsylvania, Philadelphia,
Pennsylvania 19104, U.S.A.}

\author{Maxim Sukharev}

\affiliation{Department of Physics, Arizona State University, Tempe, Arizona 85287,
USA}

\affiliation{College of Integrative Sciences and Arts, Arizona State University,
Mesa, AZ 85212, USA}

\author{Abraham Nitzan}

\affiliation{Department of Chemistry, University of Pennsylvania, Philadelphia,
Pennsylvania 19104, U.S.A.}

\author{Joseph E. Subotnik}

\affiliation{Department of Chemistry, University of Pennsylvania, Philadelphia,
Pennsylvania 19104, U.S.A.}
\begin{abstract}
In a previous paper {[}{]}, we introduced Ehrenfest+R dynamics for
a two-level system and showed how spontaneous emission can be heuristically
included such that, after averaging over an ensemble of Ehrenfest+R
trajectories, one can recover both coherent and incoherent electromagnetic
fields. In the present paper, we now show that Ehrenfest+R dynamics
can also correctly describe Raman scattering, whose features are completely
absent from standard Ehrenfest dynamics. Ehrenfest+R dynamics appear
to be quantitatively accurate both for resonant and off-resonant Raman
signals, as compared with Kramers-Heisenberg-Dirac (KHD) theory.

\end{abstract}
\maketitle

\section{Introduction\label{sec:Introduction}}

Recently there has been an explosion of interest in Raman scattering,
especially surface- and tip-enhanced Raman scattering\citep{gersten_electromagnetic_1980,weitz_enhancement_1983,willets_localized_2007,stiles_surface-enhanced_2008,morton_theoretical_2011,vasa_strong_2018},
as a probe to investigate plasmonic excitations of molecules near
a metal surface\citep{qian_surface-enhanced_2008,qian_surface_2014,masiello_many-body_2008,mirsaleh-kohan_single-molecule_2012}
and chemical reactions at catalytic surfaces.\citep{hartman_surface-_2016}
In general, the Raman technique offers the experimentalist detailed
information about how the vibrations couple to charges through electronic
polarization,\citep{hiller_can_1996,smith_toward_1996} and Raman
is very relevant for modern experiments with metallic nanoclusters.\citep{le_metallic_2008}
Furthermore, Raman spectroscopy also has the additional advantage
of offering clean signals in aqueous medium where water IR bands can
obscure signals.

From a quantitative point of view, the current theory of molecular
Raman scattering is based on the Kramers-Heisenberg-Dirac (KHD) formalism\citep{kramers_uber_1925,dirac_quantum_1927-1}
which can be reduced to Placzek's classical theory of polarizability
for off-resonance cases\citep{f._bernath_spectra_2015,tannor_introduction_2006},
as well as Albrecht's vibronic theory for resonant Raman scattering.\citep{albrecht_theory_1961,tang_developments_1970,long_raman_2002}
Over the years, efficient semiclassical tools have been developed
to evaluate Raman spectra approximately within the KHD formalsim using
an excited-state gradient approximation to propagate short time dynamics.\citep{lee_time-dependent_1979,heller_simple_1982,tannor_polyatomic_1982,heller_semiclassical_1981,tannor_introduction_2006}
More recently, chemists have also incorporated electronic structure
theories into the semiclassical description of Raman spectroscopy.\citep{neugebauer_resonance_2004,jensen_finite_2005,jensen_theory_2005,rappoport_simplified_2011}
In general, because it relies on a sum over all states (nuclear and
electronic), the KHD formalism can be difficult to implement in practice. 

One long term goal for our research groups is to study plasmonic systems
with strong light-matter couplings where Raman scattering is a very
sensitive probe of the collective behavior of electronic dynamics.\citep{sukharev_optics_2017}
For such systems, a direct implementation of KHD theory is not feasible
(because of the large number of states required) and is also likely
not relevant (because the presence of strong light-matter should invalidate
perturbation theory). Thus, in order for us to model such systems,
and to take into account strong light-matter couplings, the most natural
approach is to consider the quantum subsystems and classical electromagnetic
(EM) fields on an equal footing. This approach stands in contrast
to most existing semiclassical approaches for spectroscopy, which
treat the incoming field as an fixed external perturbation, and extrapolate
the behavior of quantum subsystems to predict light emission.\citep{lee_time-dependent_1979,heller_semiclassical_1981,milonni_semiclassical_1976,salam_molecular_2010}

Now, obviously, any computational approach to spectroscopy that promises
\textquotedbl equal footing\textquotedbl{} for light and matter will
necessarily require large approximations; in particular, we expect
that a quantum treatment of the EM field will be prohibitively difficult,
and one will necessarily need to work with classical electromagnetic
fields. The simplest example of such a mixed quantum-classical approach
is self-consistent Ehrenfest dynamics. Unfortunately, Ehrenfest dynamics
do not fully recover spontaneous emission and thus are unlikely to
capture Raman scattering either.\citep{miller_classical/semiclassical_1978,li_mixed_2018}
That being said, we are unaware of a systematic study answering this
question.

In a previous publication, our laboratory has proposed an improved
so-called \textquotedbl Ehrenfest+R\textquotedbl{} algorithm that
builds in spontaneous emission on top of Ehrenfest dynamics by enforcing
additional relaxation for two-level systems.\citep{chen_ehrenfest+r_2018}
In this second paper, our goal is to generalize Ehrenfest+R to the
case of a multi-level (i.e. more than two-level) quantum subsystem.
We will show that such a generalization can capture both resonant
and off-resonant Raman scattering (at least for a three-level molecular
system). Our results are in quantitative agreement with KHD theory.
The data presented here strongly suggests that Ehrenfest+R dynamics
(and other spruced-up versions of mean-field dynamics) can be excellent
tools for exploring interesting light-matter interactions far beyond
basic linear absorption or Raman phenomena (and also applicable to
large subsystems, e.g., plasmonic systems). 

This article is organized as follows. In Sec.~\ref{sec:II. Quantum-Theory-of-Raman},
we review the KHD formalism and calculate the polarizability and Raman
scattering profile for a three-level system. In Sec.~\ref{sec:Ehrenfest+R},
we formulate an Ehrenfest+R approach for a three-level system. In
Sec.~\ref{sec:Results}, we show Ehrenfest+R dynamics results for
Raman spectra and compare against the KHD formalism. In Sec.~\ref{sec:Conclusions},
we conclude. In this article, we use a bold symbol to denote a space
vector in Cartesian coordinate, such as $\mathbf{E}\left(\mathbf{r}\right)=E_{x}\left(\mathbf{r}\right)\hat{\mathbf{x}}+E_{y}\left(\mathbf{r}\right)\hat{\mathbf{y}}+E_{z}\left(\mathbf{r}\right)\hat{\mathbf{z}}$,
and $\widehat{H}$ denotes a quantum operator. We work in SI units.

\section{Quantum Theory of Raman Scattering\label{sec:II. Quantum-Theory-of-Raman}}

Raman light scattering is an inelastic process whereby the interaction
between the incident photons and molecules can lead to an energy shift
in emission spectra for a small fraction of the scattered photons.
To qualitatively describe Raman light scattering, consider a molecular
system with interactions between electronic states and nuclear vibrations.
Incident photons excite the molecular system to an intermediate state
(which could be a virtual state), and that intermediate state is subsequently
coupled both to the ground state as well to other vibronic states.
Thus, the system can emit photons with two different frequencies through
spontaneous emission.\citep{mukamel_principles_1999} On the one hand,
a transition back to the ground state yields scattered photons with
the same energy with the incident photons (which is known as Rayleigh
scattering). On the other hand, a transition to other vibronic states
will generate scattered photons with energies different from the incident
photons (which is known as Raman scattering). 

In this section, we review the KHD dispersion formula which quantifies
the Raman scattering cross section\citep{kramers_uber_1925,dirac_quantum_1927-1,albrecht_theory_1961}
assuming knowledge of the polarizability; we evaluate the KHD formalism
for a three-level model system in 1D space.

\subsection{Kramers-Heisenberg-Dirac Formalism}

For a quantitative description of Raman scattering, the KHD formula
is the standard, frequency domain expression for the scattering cross
section\citep{tannor_introduction_2006}

\begin{equation}
\sigma_{fi}^{\mathrm{3D}}\left(\omega_{S},\omega_{I}\right)=\frac{8\pi\omega_{I}\omega_{S}^{3}}{9c^{4}}\sum_{\rho,\lambda}\left|\left[\alpha_{fi}\left(\omega_{I}\right)\right]^{\nu\nu^{\prime}}\right|^{2},\label{eq:KHD-cross-section-1}
\end{equation}
where the polarizability is given by
\begin{equation}
\begin{split}\left[\alpha_{fi}\left(\omega_{I}\right)\right]^{\nu\nu^{\prime}}=-\sum_{k,n} & \left(\frac{\left\langle \psi_{f}\right|\widehat{\mu}^{\nu}\left|\psi_{k,n}\right\rangle \left\langle \psi_{k,n}\right|\widehat{\mu}^{\nu^{\prime}}\left|\psi_{i}\right\rangle }{\varepsilon_{i}+\hbar\omega_{I}-\varepsilon_{kn}+i\hbar\gamma}\right.\\
 & \left.+\frac{\left\langle \psi_{f}\right|\widehat{\mu}^{\nu^{\prime}}\left|\psi_{k,n}\right\rangle \left\langle \psi_{n}\right|\widehat{\mu}^{\nu}\left|\psi_{i}\right\rangle }{\varepsilon_{f}-\hbar\omega_{I}-\varepsilon_{kn}+i\hbar\gamma}\right),
\end{split}
\label{eq:KHD-polarizability-1}
\end{equation}
The frequency of the incident photons is $\omega_{I}$ and the frequency
of the scattered photons is $\omega_{S}$; these frequencies satisfy
energy conservation $\hbar\omega_{S}=\varepsilon_{i}+\hbar\omega_{I}-\varepsilon_{f}$.
The KHD formula is known as the ``sum-over-states'' formula since
the polarizability expression requires a summation over all possible
intermediate states $\psi_{k,n}$ where the index $k$ labels electronic
states and the index $n$ labels vibrational states corresponding
to electronic states. $\widehat{\mu}^{\nu}$ denotes the transition
dipole moment operator for $\nu=\left\{ x,y,z\right\} $. The linewidth
$\gamma$ corresponds to the average lifetime of the intermediate
state.\citep{jensen_theory_2005,jensen_finite_2005}

According to the scattering cross section given by Eq.~\eqref{eq:KHD-cross-section-1},
Raman spectroscopy is a two-photon spectroscopy. Experimentaly, one
typically fixes $\omega_{I}$ and observes the emission spectrum as
a function of $\omega_{S}$. The frequency $\omega_{S}=\omega_{I}$
corresponds to the contribution of Rayleigh scattering, and other
emission peaks are attributed to Raman scattering. The KHD formula
is derived using second order perturbation theory for a quantum subsystem
in the presence of the incident photons\citep{tannor_introduction_2006},
and the scattering cross section is extrapolated from the change in
electronic population.

\subsection{Three-level System}

To quantify the KHD Raman scattering formalism, we consider a model
system with three vibronic states: two lower energy states for an
electronic ground state with different vibrational states: $\left|gn_{1}\right\rangle \equiv\left|0\right\rangle $
and $\left|gn_{1}^{\prime}\right\rangle \equiv\left|1\right\rangle $
and one for an excited state $\left|en_{2}\right\rangle \equiv\left|2\right\rangle $.
We assume the energies of the vibronic states are $\varepsilon_{0}\le\varepsilon_{1}<\varepsilon_{2}$
and the electric dipole interactions couple the ground and excited
states only. Thus, the electronic Hamiltonian is time-dependent and
given by 
\begin{equation}
\widehat{H}^{\mathrm{el}}\left(t\right)=\left(\begin{array}{ccc}
\varepsilon_{0} & 0 & {\cal V}_{02}\left(t\right)\\
0 & \varepsilon_{1} & {\cal V}_{12}\left(t\right)\\
{\cal V}_{02}^{^{*}}\left(t\right) & {\cal V}_{12}^{*}\left(t\right) & \varepsilon_{2}
\end{array}\right)\label{eq:Hamiltonian}
\end{equation}
where the electric dipole coupling is 
\begin{equation}
{\cal V}_{ij}\left(t\right)=-\int\mathrm{d}x\mathbf{E}\left(x,t\right)\cdot\boldsymbol{{\cal P}}_{ij}\left(x\right).\label{eq:electric-dipole-coupling}
\end{equation}
Here we are working (without loss of generality) in 1D.

For frequency domain measurements, consider a single-mode incoming
continuous wave (CW) electromagnetic field with frequency $\omega_{I}$,
\begin{eqnarray}
\mathbf{E}_{I}\left(x,t\right) & = & \frac{A_{I}}{\sqrt{\epsilon_{0}}}\cos\left(k_{I}x-\omega_{I}t\right)\hat{\mathbf{z}}\label{eq:E_CW}\\
\mathbf{B}_{I}\left(x,t\right) & = & -\sqrt{\mu_{0}}A_{I}\sin\left(k_{I}x-\omega_{I}t\right)\hat{\mathbf{y}}\label{eq:B_CW}
\end{eqnarray}
where $\omega_{I}=ck_{I}$ and $A_{I}$ is the amplitude of the incoming
field. We assume the spatial size of the polarization is small in
space, i.e. $\boldsymbol{{\cal P}}_{ij}\left(x\right)\approx\mu_{ij}\delta\left(x\right)\hat{\mathbf{z}}$,
so that the electric dipole interactions are approximated as $\int\mathrm{d}x\mathbf{E}\left(x,t\right)\cdot\boldsymbol{{\cal P}}_{ij}\left(x\right)\approx\mu_{ij}\frac{A_{I}}{\sqrt{\epsilon_{0}}}\cos\left(\omega_{I}t\right)$. 

For light scattering in a 1D space, the scattering cross section
is defined as the ratio between the number of photons scattered per
time divided by the number of photons incident per time. With this
definition, the KHD Raman cross section becomes in 1D (see Appendix~\ref{sec:Scattering-Cross-section}):
\begin{equation}
\sigma_{fi}^{\mathrm{1D}}\left(\omega_{S},\omega_{I}\right)=\frac{\omega_{I}\omega_{S}}{2c^{2}}\left|\alpha_{fi}^{\mathrm{1D}}\left(\omega_{I}\right)\right|^{2}\label{eq:KHD-cross-section-2}
\end{equation}
For a three-level system, the KHD expression for the polarizability
for $i,f=0,1$ is
\begin{equation}
\begin{split}\alpha_{10}^{\mathrm{1D}}\left(\omega_{I}\right)=- & \left(\frac{\mu_{02}\mu_{12}}{\varepsilon_{0}+\hbar\omega_{I}-\varepsilon_{2}+i\hbar\gamma}\right.\\
 & \left.+\frac{\mu_{02}\mu_{12}}{\varepsilon_{1}-\hbar\omega_{I}-\varepsilon_{2}+i\hbar\gamma}\right).
\end{split}
\label{eq:KHD-polarizability-2}
\end{equation}
Here we take linewidth $\gamma$ to be the lifetime for the electronic
transitions of the excited state:
\begin{equation}
\frac{1}{\gamma}=\frac{1}{2}\left(\frac{1}{\kappa_{02}}+\frac{1}{\kappa_{12}}\right),\label{eq:averge_width}
\end{equation}
where the corresponding Fermi's golden rule (FGR) rates are given
by
\begin{equation}
\kappa_{fi}=\frac{\varepsilon_{i}-\varepsilon_{f}}{\hbar^{2}\epsilon_{0}c}\mu_{fi}^{2}.\label{eq:FGR_rate}
\end{equation}

In the case of resonant Raman scattering (where the incident photon
lines up with the excited state, i.e. $\varepsilon_{i}+\hbar\omega_{I}=\varepsilon_{2}$),
the first term in Eq.~\eqref{eq:KHD-polarizability-2} dominates.
Resonant Raman scattering signals are composed of two signals: (i)
When $\hbar\omega_{I}=\varepsilon_{2}-\varepsilon_{0}$ and the scattered
photon energy is $\hbar\omega_{S}=\varepsilon_{2}-\varepsilon_{1}$,
the polarizability term with $i=0$ and $f=1$ ($\alpha_{10}^{\mathrm{1D}}\left(\omega_{I}\right)$)
leads to a Stokes Raman peak (i.e. $\omega_{S}<\omega_{I}$). (ii)
When $\hbar\omega_{I}=\varepsilon_{2}-\varepsilon_{1}$ the scattered
photon energy is $\hbar\omega_{S}=\varepsilon_{2}-\varepsilon_{0}$,
the polarizability term with $i=1$ and $f=0$ ($\alpha_{01}^{\mathrm{1D}}\left(\omega_{I}\right)$)
leads to an anti-Stokes Raman peak (i.e. $\omega_{S}>\omega_{I}$).
Obviously, anti-Stokes Raman scattering can occur only state $\left|1\right\rangle $
is occupied at steady state.

In the case that the incident photon does not line up with any excited
state, the excitation is detuned far off resonance (known as off-resonance
Raman scattering). In this case, the intermediate state of the light
scattering process is a virtual state, i.e. $\varepsilon_{k}=\varepsilon_{i}+\hbar\omega_{I}$,
and the two terms in Eq.~\eqref{eq:KHD-polarizability-2} both contribute
meaningfully to the Raman cross section. Of course, for a weak field,
scattered photons are always dominated by Rayleigh scattering (i.e.
$\omega_{S}=\omega_{I}$). Note that, in the absence of pure dephasing,
there should be not fluorescence emission observed in the outgoing
field.\citep{mukamel_principles_1999}

\section{Ehrenfest+R Approach for Raman Scattering\label{sec:Ehrenfest+R}}

Given that Raman scattering is based on spontaneous emission,\citep{mukamel_principles_1999}
Ehrenfest+R dynamics should provide a proper tool for a mixed quantum-classical
simulation since the algorithm was designed to recover spontaneous
emission. One can generalize the Ehrenfest+R method to the case of
more than a two level system as follows: we add distinct +R corrections
for electronic transitions between individual pairs of states, i.e.
$2\rightarrow0$ and $2\rightarrow1$. Furthermore, to reach steady
state, we allow a phenomenological, non-radiative dissipation between
$\left|0\right\rangle $ and $\left|1\right\rangle $. In this section,
we start by formulating such a generalized Ehrenfest+R approach in
the context of the three-level system; thereafter we compare Ehrenfest+R
results against the KHD formula.

\subsection{Generalized Ehrenfest+R Method }

For the Hamiltonian given by Eq.~\eqref{eq:Hamiltonian}, there
are two electronic transitions that are mediated electric dipole couplings
${\cal V}_{02}$ and ${\cal V}_{12}$ which corresponds to spontaneous
emission rates $\kappa_{02}$ and $\kappa_{12}$ given by Eq.~\eqref{eq:FGR_rate}.
Here, based on Ref.~\onlinecite{chen_ehrenfest+r_2018}, we will
add two pairwise +R corrections on top of Ehrenfest dynamics in order
to recover the individual spontaneous emission rates ($\kappa_{fi}$)
from $\left|i\right\rangle $ to $\left|f\right\rangle $ while keeping
the other state populations fixed. 

\subsubsection{System propagator}

To implement a pairwise treatment for Ehrenfest+R dynamics, the Liouville
equation (together with additional relaxations) can be written as
\begin{equation}
\frac{\partial\widehat{\rho}}{\partial t}=-\frac{i}{\hbar}\left[\widehat{H}^{\mathrm{el}},\widehat{\rho}\right]+\doublehat{{\cal L}_{\hspace{0.1pt}}}_{R}^{2\rightarrow0}\widehat{\rho}+\doublehat{{\cal L}_{\hspace{0.1pt}}}_{R}^{2\rightarrow1}\widehat{\rho},\label{eq:Liouville+R}
\end{equation}
Here, the diagonal elements of the $\doublehat{{\cal L}_{\hspace{0.1pt}}}_{R}^{i\rightarrow f}$
super-operators are defined by 
\begin{equation}
\left[\doublehat{{\cal L}_{\hspace{0.1pt}}}_{R}^{i\rightarrow f}\widehat{\rho}\right]_{ii}=-\left[\doublehat{{\cal L}_{\hspace{0.1pt}}}_{R}^{i\rightarrow f}\widehat{\rho}\right]_{ff}=-k_{R}^{fi}\rho_{ii},\label{eq:Liouville+R_diagonal}
\end{equation}
and the off-diagonal element of $\left[{\cal L}_{R}\widehat{\rho}\right]_{ij}$
are chosen to be
\begin{equation}
\left[\doublehat{{\cal L}_{\hspace{0.1pt}}}_{R}^{i\rightarrow f}\widehat{\rho}\right]_{if}=\left[\doublehat{{\cal L}_{\hspace{0.1pt}}}_{R}^{i\rightarrow f}\widehat{\rho}\right]_{fi}^{*}=-\gamma_{R}^{fi}\rho_{if}.\label{eq:Liouville+R_offdiagonal}
\end{equation}
 The +R relaxation rate $k_{R}^{fi}$ for the transistion $i\rightarrow f$
is given by
\begin{equation}
k_{R}^{fi}\equiv2\kappa_{fi}\left(1-\rho_{ff}\right)\text{Im}\left[\frac{\rho_{fi}}{\left|\rho_{fi}\right|}e^{i\phi}\right]^{2}.\label{eq:k_R}
\end{equation}
Here, the $\kappa_{fi}$ is the FGR in Eq~\eqref{eq:FGR_rate}.
$\phi\in\left(0,2\pi\right)$ is a phase chosen randomly for each
Ehrenfest+R trajectory. The +R dephasing rate $\gamma_{R}^{fi}$ in
Eq.~\eqref{eq:Liouville+R_offdiagonal} is chosen to be
\begin{equation}
\gamma_{R}^{fi}\equiv\frac{\kappa_{fi}}{2}\left(1-\rho_{ff}+\rho_{ii}\right).\label{eq:gamma_R}
\end{equation}

In practice, we use a pure state representation for the density matrix:
$\widehat{\rho}=\left|\psi\right\rangle \left\langle \psi\right|$
with wavefunction $\left|\psi\left(t\right)\right\rangle =c_{0}\left(t\right)\left|0\right\rangle +c_{1}\left(t\right)\left|1\right\rangle +c_{2}\left(t\right)\left|2\right\rangle $.
The additional relaxation embodied by $\doublehat{{\cal L}_{\hspace{0.1pt}}}_{R}^{i\rightarrow f}$
is defined by a transition operator:
\begin{equation}
\widehat{{\cal T}}\left[k_{R}^{fi}\right]\left(\begin{array}{c}
\vdots\\
c_{i}\\
\vdots\\
c_{f}\\
\vdots
\end{array}\right)=\left(\begin{array}{c}
\vdots\\
\frac{c_{i}}{\left|c_{i}\right|}\sqrt{\left|c_{i}\right|^{2}-k_{R}^{fi}\left|c_{i}\right|^{2}dt}\\
\vdots\\
\frac{c_{f}}{\left|c_{f}\right|}\sqrt{\left|c_{f}\right|^{2}+k_{R}^{fi}\left|c_{f}\right|^{2}dt}\\
\vdots
\end{array}\right),\label{eq:transistion_operator}
\end{equation}
with a fixed relative phase between $c_{i}$ and $c_{f}$, plus a
stochastic random phase operator:
\begin{equation}
e^{i\widehat{\Phi}\left[\gamma_{R}^{fi}\right]}\left(\begin{array}{c}
\vdots\\
c_{i}\\
\vdots\\
c_{f}\\
\vdots
\end{array}\right)=\left(\begin{array}{c}
\vdots\\
e^{i\Phi_{i}}c_{i}\\
\vdots\\
e^{i\Phi_{f}}c_{f}\\
\vdots
\end{array}\right)\ \text{if RN}<\gamma_{R}^{fi}dt.\label{eq:random_phase_operator}
\end{equation}
Here $\text{RN}\in\left[0,1\right]$ is a random number, and we choose
$\Phi_{i}=0$, $\Phi_{f}\in\left(0,2\pi\right)$ as random phases.
In other words, we choose to give a random phase only to the final
state ($f$) which has a lower energy than the initial state ($i$).
This choice is crucial for ensuring that, e.g., spontaneous emission
from $2\rightarrow1$ does not affect the coherence between states
$2$ and $0$.

\begin{widetext}Thus in practice, the time evolution of the subsystem
wavefunction is carried out as
\begin{equation}
\left|\psi\left(t+dt\right)\right\rangle =e^{i\widehat{\Phi}\left[\gamma_{R}^{12}\right]}\widehat{{\cal T}}\left[k_{R}^{12}\right]\times e^{i\widehat{\Phi}\left[\gamma_{R}^{02}\right]}\widehat{{\cal T}}\left[k_{R}^{02}\right]\times e^{-i\widehat{H}^{\mathrm{el}}dt/\hbar}\left|\psi\left(t\right)\right\rangle .\label{eq:propagation-system}
\end{equation}
\end{widetext}where $e^{-i\widehat{H}^{\mathrm{el}}dt/\hbar}$ is
responsible for propagating according to the first term of Eq.~\eqref{eq:Liouville+R}.
Note that $e^{i\widehat{\Phi}\left[\gamma_{R}^{12}\right]}\widehat{{\cal T}}\left[k_{R}^{12}\right]$
and $e^{i\widehat{\Phi}\left[\gamma_{R}^{02}\right]}\widehat{{\cal T}}\left[k_{R}^{02}\right]$
commute as long as $dt$ is sufficiently small. 

\subsubsection{EM field propagator}

We write the total EM field in the form of $\mathbf{E}=\mathbf{E}_{I}+\mathbf{E}_{S}$
and $\mathbf{B}=\mathbf{B}_{I}+\mathbf{B}_{S}$ where $\mathbf{E}_{S}$
and $\mathbf{B}_{S}$ are the scattered EM fields. For a CW field
given by Eq.~\eqref{eq:E_CW} and Eq.~\eqref{eq:B_CW}, $\mathbf{E}_{I}$
and $\mathbf{B}_{I}$ satisfy source-less Maxwell's equations, so
we can treat the CW field as a standalone external field. Therefore,
for underlying Ehrenfest dynamics, the scattered fields $\mathbf{E}_{S}$
and $\mathbf{B}_{S}$ satisfy Maxwell's equations:
\begin{eqnarray}
\frac{\partial}{\partial t}\mathbf{B}_{S} & = & -\boldsymbol{\nabla}\times\mathbf{E}_{S},\label{eq:maxwell_BE}\\
\frac{\partial}{\partial t}\mathbf{E}_{S} & = & c^{2}\boldsymbol{\nabla}\times\mathbf{B}_{S}-\frac{1}{\epsilon_{0}}\mathbf{J},\label{eq:maxwell_EB}
\end{eqnarray}
where the average current is
\begin{equation}
\mathbf{J}\left(x,t\right)=\sum_{i=2}\sum_{f=0,1}2\left(\varepsilon_{f}-\varepsilon_{i}\right)\mathrm{Im}\left[\rho_{fi}\left(t\right)\right]\boldsymbol{{\cal P}}_{fi}\left(x\right).
\end{equation}

Given the pairwise transitions of the subsystem, the classical EM
field must be rescaled. We denote the rescaling operator for the
EM fields by:
\begin{equation}
{\cal R}\left[\delta U_{R}^{fi}\right]:\left(\begin{array}{c}
\mathbf{E}_{S}\\
\mathbf{B}_{S}
\end{array}\right)\rightarrow\left(\begin{array}{c}
\mathbf{E}_{S}+\alpha^{fi}\delta\mathbf{E}_{R}^{fi}\\
\mathbf{B}_{S}+\beta^{fi}\delta\mathbf{B}_{R}^{fi}
\end{array}\right).\label{eq:rescaling_operator}
\end{equation}
where the rescaling coefficients are chosen to be
\begin{eqnarray}
\text{\ensuremath{\alpha^{fi}}} & = & \sqrt{\frac{cdt}{\Lambda^{fi}}\frac{\delta U_{R}^{fi}}{\epsilon_{0}\int\mathrm{d}v\left|\delta\mathbf{E}_{R}^{fi}\right|^{2}}}\times\text{sgn}\left(\text{Im}\left[\rho_{fi}e^{i\phi}\right]\right),\\
\text{\ensuremath{\beta^{fi}}} & = & \sqrt{\frac{cdt}{\Lambda^{fi}}\frac{\mu_{0}\delta U_{R}^{fi}}{\int\mathrm{d}v\left|\delta\mathbf{B}_{R}^{fi}\right|^{2}}}\times\text{sgn}\left(\text{Im}\left[\rho_{fi}e^{i\phi}\right]\right).
\end{eqnarray}
Here $\Lambda^{fi}$ is the self-interference length (see Ref.~\onlinecite{chen_ehrenfest+r_2018}).
For a Gaussian polarization profile (as in Eq.~\eqref{eq:P_1D})
$\Lambda^{fi}=2.363/\sqrt{2a}$. The energy change for each pairwise
relaxation $i\rightarrow f$ is 
\begin{equation}
\delta U_{R}^{fi}=\left(\varepsilon_{i}-\varepsilon_{f}\right)k_{R}^{fi}\rho_{ii}dt.\label{eq:EnergyChange}
\end{equation}
According to Eq.~\eqref{eq:propagation-system}, we need to perform
two rescaling operators (${\cal R}\left[\delta U_{R}^{12}\right]$
and ${\cal R}\left[\delta U_{R}^{02}\right]$) corresponding to the
two relaxation pathways ($2\rightarrow0$ and $2\rightarrow1$).

For the results below, we assume that the transition dipole moments
are the same for both the $2\rightarrow1$ and $2\rightarrow0$ transitions,
i.e.$\boldsymbol{{\cal P}}_{02}=\boldsymbol{{\cal P}}_{12}=\boldsymbol{{\cal P}}$,
so that the rescaling fields can be chosen to be $\delta\mathbf{E}_{R}^{fi}=\delta\mathbf{E}_{R}$
and $\delta\mathbf{B}_{R}^{fi}=\delta\mathbf{B}_{R}$. For a 1D system,
the rescaling fields take the form
\begin{eqnarray}
\delta\mathbf{E}_{R} & = & \boldsymbol{\nabla}\times\boldsymbol{\nabla}\times\boldsymbol{{\cal P}}-g\boldsymbol{{\cal P}},\label{eq:rescaling_E}\\
\delta\mathbf{B}_{R} & = & -\boldsymbol{\nabla}\times\boldsymbol{{\cal P}}-h\left(\boldsymbol{\nabla}\times\right)^{3}\boldsymbol{{\cal P}},\label{eq:rescaling_B}
\end{eqnarray}
As demonstrated in Ref.~\onlinecite{chen_ehrenfest+r_2018}, for
Gaussian polarization, we choose $g=2a$ and $h=1/6a$. With this
assumption, we can combine the two rescaling operators as ${\cal R}\left[\delta U_{R}^{12}+\delta U_{R}^{02}\right]$.

In the end, each Ehrenfest+R trajectory for classical EM fields is
propagated by 
\begin{equation}
\left(\begin{array}{c}
\mathbf{E}_{S}\left(t+dt\right)\\
\mathbf{B}_{S}\left(t+dt\right)
\end{array}\right)={\cal R}\left[\delta U_{R}^{12}+\delta U_{R}^{02}\right]{\cal M}\left[dt\right]\left(\begin{array}{c}
\mathbf{E}_{S}\left(t\right)\\
\mathbf{B}_{S}\left(t\right)
\end{array}\right).\label{eq:propagation-field}
\end{equation}
Here ${\cal M}\left[dt\right]$ denotes the linear propagator of Maxwell's
equations (Eq.~\eqref{eq:maxwell_BE} and \eqref{eq:maxwell_EB})
for time step $dt$. 

\subsubsection{Non-radiative dissipation}

Without any dissipation allowed, the three-level system in Eq.~\eqref{eq:Hamiltonian}
eventually reaches the asymptotic state $\left|\psi\left(t\rightarrow\infty\right)\right\rangle =\left|1\right\rangle $
in the presence of the CW field. By contrast, to reach the correct
steady state, we must take into account vibrational relaxation. Thus,
we also introduce a phenomenological, non-radiative relaxation from
$\left|1\right\rangle $ to $\left|0\right\rangle $ by a transition
operator:
\begin{equation}
\left|\psi\left(t+dt\right)\right\rangle \rightarrow\widehat{{\cal T}}\left[k_{\mathrm{vib}}^{01}\right]\left|\psi\left(t+dt\right)\right\rangle .\label{eq:propagation-system-1}
\end{equation}
where the operation of the transition operator ${\cal T}$ is defined
in Eq.~\eqref{eq:transistion_operator}. The classical EM field is
not rescaled for this non-radiative transition, and the vibrational
decay rate $k_{\mathrm{vib}}^{01}$ is an empirical parameter which
will to be specified later. Note that we exclude thermal transitions
from $\left|0\right\rangle $ to $\left|1\right\rangle $ since we
assume the system is at a very low temperature.

In the end, Ehrenfest+R dynamics are specified by Eqs.~\eqref{eq:propagation-system},
\eqref{eq:propagation-field}, and \eqref{eq:propagation-system-1}.

\subsubsection{Coherent and incoherent emission}

Our primary interest is in the scattering EM field when the system
reaches steady state ($t\rightarrow t_{\mathrm{ss}}$) in presence
of an external CW field. Let $\left\{ \ensuremath{\mathbf{E}_{S}^{\ell}\left(x,t_{\mathrm{ss}}\right)};\ell\in N_{\mathrm{traj}}\right\} $
be the set of scattering electric fields at a steady state for an
ensemble of Ehrenfest+R trajectories (labeled by $\ell$). The average
electric field $\left\langle \mathbf{E}_{S}\left(x,t_{\mathrm{ss}}\right)\right\rangle $
represents coherent emission, and the Fourier transform of the average
electric field yields the scattering spectrum for coherent emission:
\begin{equation}
\left\langle \mathbf{E}_{S}\left(\omega_{S}\right)\right\rangle =\int dxe^{i\omega_{S}x/c}\frac{1}{N_{\mathrm{traj}}}\sum_{\ell}^{N_{\mathrm{traj}}}\text{\ensuremath{\mathbf{E}_{S}^{\ell}\left(x,t_{\mathrm{ss}}\right)}}.\label{eq:average_E}
\end{equation}
We expect that, in Eq.~\eqref{eq:average_E}, all incoherent contributionswith
random phases will vanish when we take ensemble average. We denote
the magnitude of the coherent emission intensity at scattering frequency
$\omega_{S}$ as $\left|\left\langle \mathbf{E}_{S}\left(\omega_{S}\right)\right\rangle \right|^{2}$.

We now turn to the incoherent emission. The expectation value of the
intensity distribution $\left\langle \left|\mathbf{E}_{S}\left(x,t_{\mathrm{ss}}\right)\right|^{2}\right\rangle $
corresponds to the energy distribution of the scattering EM field.
We can obtain the total emission power spectrum by averaging over
the intensity in Fourier space:
\begin{equation}
\left\langle \left|\mathbf{E}_{S}\left(\omega_{S}\right)\right|^{2}\right\rangle =\frac{1}{N_{\mathrm{traj}}}\sum_{\ell}^{N_{\mathrm{traj}}}\text{\ensuremath{\left|\int dxe^{i\omega_{S}x/c}\mathbf{E}_{S}^{\ell}\left(x,t_{\mathrm{ss}}\right)\right|^{2}}}.\label{eq:E_average}
\end{equation}
Note that the total intensity in Eq.~\eqref{eq:E_average} includes
the contributions of both coherent and incoherent scattering signals.
Thus, $\left\langle \left|\mathbf{E}_{S}\left(\omega_{S}\right)\right|^{2}\right\rangle $
can be considered as the energy distribution of scattering photons
with mode $\omega_{S}$. Finally we can extract a scattering cross
section from Ehrenfest+R dynamics by the formula
\begin{equation}
\sigma_{10}^{\mathrm{1D}}\left(\omega_{S},\omega_{I}\right)=\frac{\left\langle \left|\mathbf{E}_{S}\left(\omega_{S}\right)\right|^{2}\right\rangle /\omega_{S}}{A_{I}^{2}/\omega_{I}},\label{eq:cross_section_1D}
\end{equation}
according to the definition of 1D scattering cross section and the
Einstein relation (see Eq.~\ref{eq:Einstein-photon}). 

Before concluding this section, let us once more emphasize the obvious
conclusion of Ref.~\onlinecite{chen_ehrenfest+r_2018}. Within Ehrenfest+R
dynamics, standard Ehrenfest dynamics yields only coherent emission;
at the same time, however, the +R relaxation pathway is able to produce
incoherent emission.

\section{Results\label{sec:Results}}

\subsection{Parameters}

As far as simulating Raman scattering by Ehrenfest+R approach, we
consider a three-level system with $\varepsilon_{0}=0$, $\varepsilon_{1}=4.115\ \mathrm{eV}$,
and $\varepsilon_{2}=16.46\ \mathrm{eV}$, so that we can define the
frequencies of the system: $\hbar\Omega_{20}=\varepsilon_{2}-\varepsilon_{0}$
and $\hbar\Omega_{21}=\varepsilon_{2}-\varepsilon_{1}$. For convenience,
we let $\Omega_{20}=\Omega$ and $\Omega_{21}=\frac{3}{4}\Omega$.
We assume the initial state of the system is the ground state, $\left|\psi\left(t=0\right)\right\rangle =\left|0\right\rangle $
and we turn on the incident CW field at $t=0$. The transition dipole
moment takes the form of a Gaussian distribution:
\begin{equation}
\boldsymbol{{\cal P}}_{02}\left(x\right)=\boldsymbol{{\cal P}}_{12}\left(x\right)=\mu\sqrt{\frac{a}{\pi}}e^{-ax^{2}}\hat{\mathbf{z}},\label{eq:P_1D}
\end{equation}
where $\mu=11282\ \text{C/nm/mol}$ and $a=1/2\sigma^{2}$ with $\sigma=3.0\ \text{nm}$.
With this polarization, the rescaling fields are (from Ref.~\onlinecite{chen_ehrenfest+r_2018}):
\begin{eqnarray}
\delta\mathbf{E}_{R}\left(x\right) & = & -\mu\sqrt{\frac{a}{\pi}}4a^{2}x^{2}e^{-ax^{2}}\hat{\mathbf{z}},\label{eq:dE_1D-1}\\
\delta\mathbf{B}_{R}\left(x\right) & = & \mu\sqrt{\frac{a}{\pi}}\frac{4}{3}a^{2}x^{3}e^{-ax^{2}}\hat{\mathbf{y}}.\label{eq:dB_1D-1}
\end{eqnarray}
The average lifetime is $1/\gamma\approx40\ \mathrm{fs}$. We run
dynamics for $t_{\mathrm{ss}}=200\ \text{fs}$ to reach a steady state,
averaging over $N_{\mathrm{traj}}=400$ trajectories. For the non-radiative
dissipation, we choose a vibrational decay rate to be $k_{\mathrm{vib}}^{01}/\gamma=37.33$.
Note that, as long as $k_{\mathrm{vib}}^{01}\gg\gamma$ is large enough,
our results do not depend on the choice of $k_{\mathrm{vib}}^{01}$.

\begin{figure*}
\noindent \begin{centering}
\includegraphics{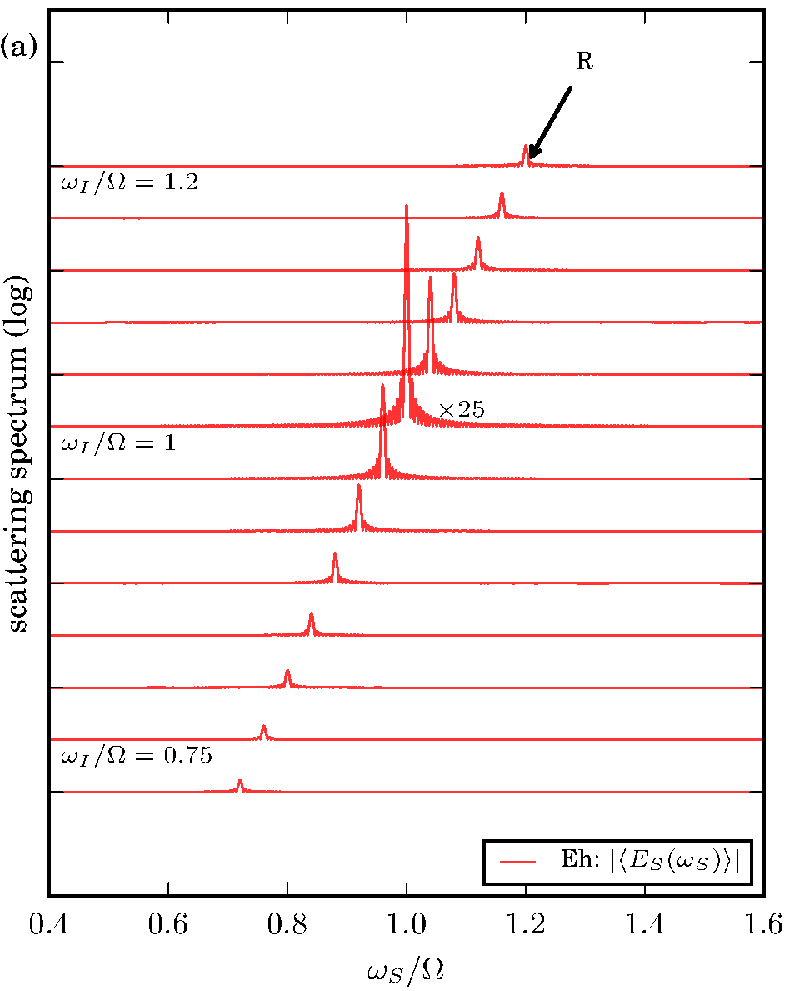}\includegraphics{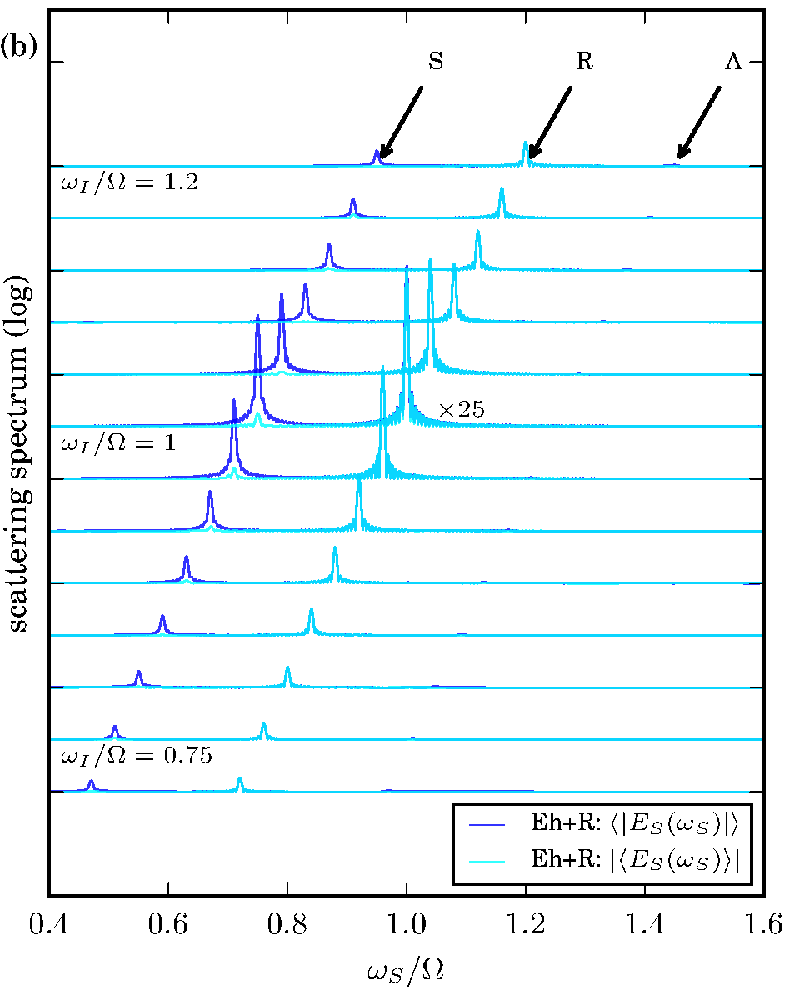}
\par\end{centering}
\noindent \begin{centering}
\includegraphics{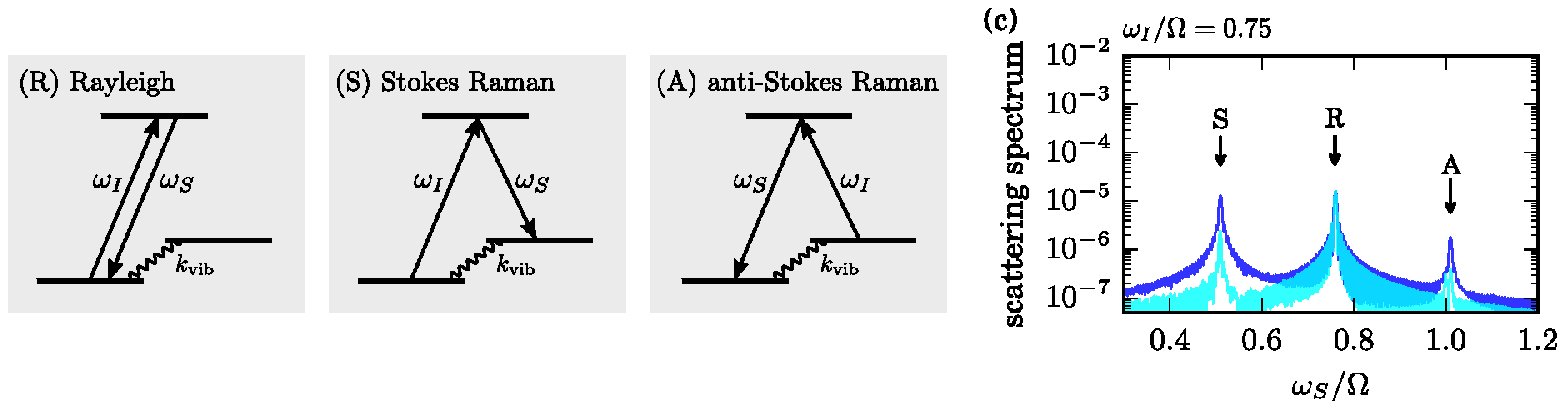}
\par\end{centering}
\caption{Raman scattering spectra as a function of $\omega_{S}/\Omega$ when
varying the incident CW field frequency $\omega_{I}/\Omega$. We plot
the total intensity spectrum $\left\langle \left|\mathbf{E}_{S}\left(\omega_{S}\right)\right|^{2}\right\rangle ^{1/2}=\left|\left\langle \mathbf{E}_{S}\left(\omega_{S}\right)\right\rangle \right|$
obtained by standard Ehrenfest dynamics in (a). For Ehrenfest+R dynamics,
we plot both the coherent emission spectrum $\left|\left\langle \mathbf{E}_{S}\left(\omega_{S}\right)\right\rangle \right|$
(colored cyan) and the total intensity spectrum $\left\langle \left|\mathbf{E}_{S}\left(\omega_{S}\right)\right|^{2}\right\rangle ^{1/2}$
(colored blue). The incoming field amplitude is $A_{I}/\sqrt{\hbar\Omega}=6\times10^{-3}$.
For all CW frequencies, Rayleigh scattering peaks are observed at
$\omega_{S}=\omega_{I}$. Stokes Raman scattering is always observed
at $\omega_{S}=\omega_{I}-\frac{1}{4}\Omega$. In the case of resonant
Raman, when $\omega_{I}/\Omega=1$, a strong Stokes signal occurs
at $\omega_{S}/\Omega=\frac{3}{4}$; there is also a small anti-Stokes
signal occurring at $\omega_{S}/\Omega=1$ when $\omega_{I}/\Omega=\frac{3}{4}$
. Obviously, the anti-Stokes resonant Raman signal is always much
smaller than the Stokes Raman signal, on or off resonance. (c) A semi-log
plot of the scattering spectrum for $\omega_{I}/\Omega=\frac{3}{4}$.
With this log scale, one can clearly see that Ehrenfest+R dynamics
recovers both Stokes and anti-Stokes Raman scattering peaks (whereas
standard Ehrenfest dynamics produces only Rayleigh scattering). Note
also that only Rayleigh scattering comes in the form of a cohrerent
emission field; Raman scattering are both incoherent emission fields.\label{fig:Raman-spectra-with}}
\end{figure*}

\subsubsection{Resonance and off-resonance scattering}

We first focus on Raman scattering in the frequency-domain spectrum.
In Fig.~\ref{fig:Raman-spectra-with}, we plot the spectrum of coherent
emission and total scattering at steady state as a function of $\omega_{S}$
for various incident frequencies $\omega_{I}$. In Fig.~\ref{fig:Raman-spectra-with}(a),
we plot results from Ehrenfest dynamics, and in Fig.~\ref{fig:Raman-spectra-with}(b),
we plot results from Ehrenfest+R dynamics. When the incident field
is far from resonance, we find that the scattered EM field is dominated
by Rayleigh scattering ($\omega_{S}=\omega_{I}$), as expected from
the KHD formula. Qualitatively, both standard Ehrenfest dynamics and
Ehrenfest+R dynamics predict Rayleigh scattering peaks at the correct
frequency and show a linear shift with respect to the incident frequency.
When the incident photon is at resonance (i.e. the incident frequency
$\omega_{I}$ lines up with electronic excitation), Ehrenfest+R dynamics
captures Raman scattering peaks at $\left(\omega_{I},\omega_{S}\right)=\text{\ensuremath{\left(\Omega_{20},\Omega_{21}\right)}}$
qualitatively. Note that anti-Stokes Raman scattering is relatively
weak here. 

In contrast to Ehrenfest+R dynamics, we also plot the spectra obtained
from standard Ehrenfest calculations in Fig.\ref{fig:Raman-spectra-with}(a).
From Fig.~\ref{fig:Raman-spectra-with}, we must emphasize that Ehrenfest
dynamics capture only Rayleigh scattering peaks, but not Raman scattering
peaks. To rationalize this behavior, we recall that the Ehrenfest
decay rate for spontaneous emission depends linearly on the lower
state population.\citep{chen_ehrenfest+r_2018} For the initial state
$c_{0}=1$ and $c_{1}=c_{2}=0$, the system is excited to state $\left|2\right\rangle $
by the incident field, but will never populate state $\left|1\right\rangle $.
Therefore, effectively we always have $c_{1}=0$ within Ehrenfest
dynamics and the spontaneous emission via electronic transition $2\rightarrow1$
never occurs. As a general rule of thumb, because standard Ehrenfest
dynamics are effectively classical dynamics, whereas there is only
a single frequency $\omega_{I}$ in the EM field and the EM field
strength is weak, Ehrenfest dynamics will predict all response to
be at the same frequency $\omega_{I}$.

\subsubsection{Coherent emission and total intensity}

Several words are now appropriate regarding the character of the outgoing
fields: are they coherent (with $\left|\left\langle \mathbf{E}_{S}\right\rangle \right|^{2}=\left\langle \left|\mathbf{E}_{S}\right|^{2}\right\rangle $),
are they partially coherent, or are they totally incoherent? In Fig.~\ref{fig:Raman-spectra-with}(b),
we observe that Rayleigh scattering is made up of completely coherent
emission according to Ehrenfest+R dynamics. For an elastic scattering
process ($\omega_{S}=\omega_{I}$) such as Rayleigh scattering, the
outgoing field retains the phase of the incoming field, so that the
signal is not canceled out in the average electric field $\left\langle \mathbf{E}_{S}\left(\omega_{S}\right)\right\rangle $. 

By contrast, Raman scattering peaks are dominated by incoherent
signals. For these signals, the coherent emission is much smaller
than total scattering intensity, i.e. $\left|\left\langle \mathbf{E}_{S}\left(\omega_{S}\right)\right\rangle \right|^{2}\ll\left\langle \left|\mathbf{E}_{S}\left(\omega_{S}\right)\right|^{2}\right\rangle $.
To understand this, we show in Appendix \ref{sec:Coherent-emission-intensity}
that, for a simplified model within the rotating wave approximation,
the average electric field does not include a contribution at frequency
$\omega_{S}=\Omega_{21}$. Instead, the signal is incoherent, as the
coherence of the incoming EM field is disturbed by the Raman inelastic
light scattering process. Note that, within Ehrenfest+R dynamics,
this incoherence is introduced by applying the stochastic random phase
operators in Eq.~\eqref{eq:random_phase_operator}.

\begin{figure}[t]
\noindent \begin{centering}
\includegraphics{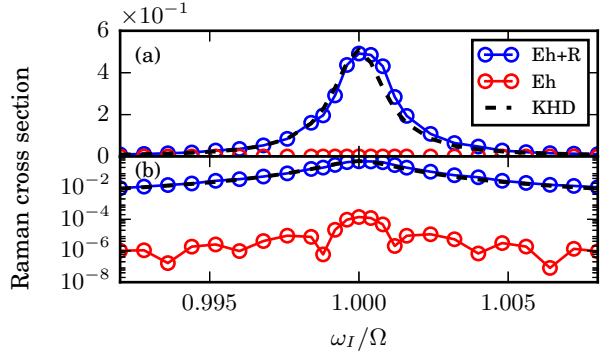}
\par\end{centering}
\caption{The Raman scattering cross section as a function of incident frequency
near resonance ($\omega_{I}/\Omega\approx1$). Standard Ehrenfest
dynamics are colored red, and Ehrenfest+R dynamics are colored blue.
The KHD formula is plotted in dashed line. The incoming field amplitude
is $A_{I}/\sqrt{\hbar\Omega}=6\times10^{-3}$. (a) is a linear plot
and (b) is a semi-log plot. Note that Ehrenfest+R dynamics match the
KHD Raman signal, whereas Ehrenfest dynamics alone do not. \label{fig:3LS_KCW} }
\end{figure}

\subsubsection{Resonant Raman cross section }

We now turn our attention to the near-resonant regime, i.e. $\omega_{I}\approx\Omega_{20}$,
and focus on Raman scattering. To compare against the KHD formula,
we extract the scattering cross section from Ehrenfest+R dynamics
by Eq.~\eqref{eq:cross_section_1D}.In Fig.~\ref{fig:3LS_KCW}(a),
we compare Ehrenfest+R dynamics with the KHD formula (Eq.~\eqref{eq:KHD-cross-section-2}).
We demonstrate that Ehrenfest+R dynamics can quantitatively recover
the enhancement of the Raman scattering cross section in the nearly
resonant regime, while the standard Ehrenfest dynamics does not predict
any enhancement. Furthermore, the linewidth obtained by Ehrenfest+R
approach agrees with the average lifetime for the KHD formula (Eq.~\eqref{eq:averge_width}).
In Fig.~\ref{fig:3LS_KCW}(b), the difference between standard Ehrenfest
and Ehrenfest+R results are plotted in logarithmic scale. The +R correction
is necessary in order for semiclassical simulations to recover resonance
Raman scattering. 

\begin{figure}
\noindent \begin{centering}
\includegraphics[bb=0bp 0bp 227bp 193bp]{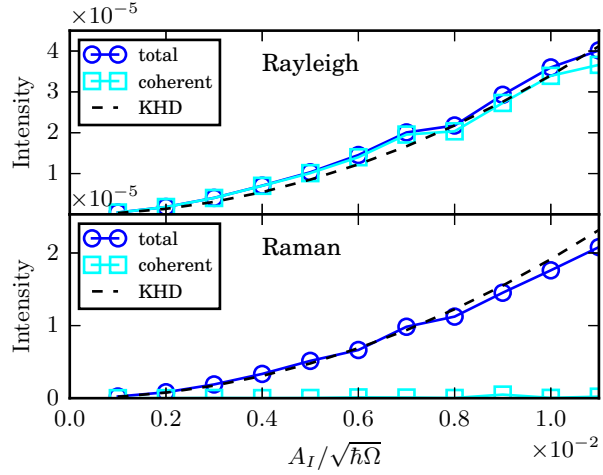}
\par\end{centering}
\caption{The resonant scattering intensity as a function of incident CW amplitude
$A_{I}$. The incident frequency is $\omega_{I}=\Omega_{20}$. The
upper panel is the Rayleigh signal ($\omega_{S}=\Omega$) and the
lower panel is the Raman signal ($\omega_{S}=\Omega_{21}$). For Ehrenfest+R
dynamics, the blue circles represent the total intensity ($\left\langle \left|E_{S}\left(\omega_{S}\right)\right|^{2}\right\rangle $)
and the green squares represent the coherent emission intensity ($\left|\left\langle E_{S}\left(\omega_{S}\right)\right\rangle \right|^{2}$).
The black dashed line is the total intensity given by the KHD formula
(Eq.~\eqref{eq:total_intensity}). The value of the KHD intensity
in the lower panel is exactly $\left(\Omega_{21}/\Omega_{20}\right)^{2}=\left(3/4\right)^{2}$
of the upper panel. Note that both KHD and Ehrenfest+R correctly capture
the Raman and Rayleigh signals that are linear with respect to the
incoming filed. Note also that the Raman signal is incoherent, whereas
the Rayleigh signal is almost entirely coherent.\label{fig:3LS_ACW} }
\end{figure}

\subsubsection{Field strength}

Finally we focus on the intensity of resonance Raman scattering (i.e.
$\omega_{I}=\Omega_{20}$) in response to various incident field amplitudes.
Indeed, one might question whether or not the Raman signals as predicted
by Ehrefenst+R dynamics scale correctly with respect to the EM field
strength; indeed a devil's advocate might argue that these ``Raman-like''
feature emerging from semiclassical dynamics are really non-linear
features that arise from strong EM fields incident in the molecule.
And yet, it is crucial to emphasize that Raman is a linear spectroscopy.
From the KHD formula, the resonant scattering signal intensity in
the weak field regime scales as $\left|\mathbf{E}_{S}\right|\sim A_{I}$
for all scattered frequencies $\omega_{S}$:
\begin{equation}
\left\langle \left|\mathbf{E}_{S}\left(\omega_{S}\right)\right|^{2}\right\rangle =A_{I}^{2}\frac{\omega_{S}^{2}}{2\hbar^{2}c^{2}}\frac{\mu^{4}}{\gamma^{2}}.\label{eq:total_intensity}
\end{equation}
Here $\gamma$ is given in Eq.~\eqref{eq:averge_width}. Furthermore,
we note that, from Eq.~\eqref{eq:total_intensity}, one can derive
a simple relation for the ratio of the intensity for Raman scattering
($\hbar\omega_{S}=\Omega_{21}$) and for Rayleigh scattering ($\hbar\omega_{S}=\Omega_{20}$)
given by
\begin{equation}
\frac{\left\langle \left|\mathbf{E}_{S}\left(\omega_{S}=\Omega_{21}\right)\right|^{2}\right\rangle }{\left\langle \left|\mathbf{E}_{S}\left(\omega_{S}=\Omega_{20}\right)\right|^{2}\right\rangle }=\left(\frac{\Omega_{21}}{\Omega_{20}}\right)^{2}.\label{eq:intensity_ratio}
\end{equation}
Do Ehrenfest+R dynamics capture these scaling relationships? To answer
these questions, in Fig~\ref{fig:3LS_ACW} we plot the Raman and
Rayleigh scattering intensity signals as obtained from Ehrenfest+R
dynamics as a function of $A_{I}$. We show conclusively that the
Ehrenfest+R signals is linear with respect to $A_{I}$, in agreement
with the KHD formula. This also shows that the ratio of the Raman
and Rayleigh signals agrees with Eq.~\eqref{eq:intensity_ratio}.

To contrast the coherent emission with the total scattering intensity,
we also plot the coherent emission intensity ($\left|\left\langle E_{S}\left(\omega_{S}\right)\right\rangle \right|^{2}$)
at the Raman and Rayleigh frequencies as a function of $A_{I}$. As
we discussed above, the coherent emission of Raman scattering is approximately
zero for all $A_{I}$. By contrast, the signal at frequency $\omega_{S}=\omega_{I}=\Omega_{20}$
is almost exclusively a coherent Rayleigh scattering signal.

\section{Discussion and Conclusions\label{sec:Conclusions}}

In this work, we have generalized the Ehrenfest+R approach to treat
a multi-level (more than two-level) system and we have demonstrated
that such an approach recapitulates Raman scattering. In the context
of a three-level system model, the proposed prescription of +R corrections
can overcome the qualitative deficiencies of Ehrenfest dynamics and
recover both resonant and off-resonant Raman scattering. In addition,
a comparison with the quantum mechanical KHD formalism shows that
Ehrenfest+R dynamics agrees quantitatively with resonant Raman scattering
cross sections.

Given the promising results in this work, there are many further questions
that need to be addressed. First, the proposed prescription is based
on pairwise +R transitions with stochastic random phases for decoherence.
If we take into account pure dephasing of the the system, can this
prescription of Ehrenfest+R dynamics produce the correct (and fully
incoherent) fluorescence signals? More generally, have we found the
optimal semiclassical approach for quantum electrodynamics with more
than two electronic states? It will be very interesting to compare
the present Ehrenfest+R approach with more standard nonadiabatic approaches,
including PLDM,\citep{huo_consistent_2012} PBME,\citep{kim_quantum-classical_2008}
and SQC\citep{miller_classical/semiclassical_1978} (which has shown
great promise for spin-boson Hamiltonians). Second, the data in this
work was generated for a three level system in one dimension only,
assuming that the polarization density has a simple Gaussian profile.
Does our prescription work for a system with arbitrary polarization
density in three dimensions? Finally, the current setup includes one
quantum subsystem only. How can we to treat the collective behavior
of a set of molecular subsystems with strong electronic coupling?
These questions will be investigated in the future.

\section*{Acknowledgment}

J.E.S. acknowledges start up funding from the University of Pennsylvania.
The research of AN is supported by the Israel-U.S. Binational Science
Foundation, the German Research Foundation (DFG TH 820/11-1), the
U.S. National Science Foundation (Grant No. CHE1665291), and the University
of Pennsylvania. M.S. would also like to acknowledge financial support
by the Air Force Office of Scientific Research under Grant No. FA9550-15-1-0189
and Binational Science Foundation under Grant No. 2014113. 

\appendix

\section{Scattering Cross section in a 1D space \label{sec:Scattering-Cross-section}}

Here we derive the scattering cross section for a 1D system within
the KHD formalism. Following Tannor's approach in Ref.~\onlinecite{tannor_introduction_2006},
we make the rotating wave approximation (RWA) such that the electric
dipole coupling can be written as $\widehat{\mu}E_{I}e^{-i\omega_{I}t}/2$
for an incoming photon with amplitude $E_{I}$ and frequency $\omega_{I}$,
and $\widehat{\mu}E_{S}e^{i\omega_{S}t}/2$ for an outgoing photon
with amplitude $E_{S}$ and frequency $\omega_{S}$. (The amplitude
$E_{S}$ will be determined below.) Here $\widehat{\mu}$ is the dipole
operator of the electronic system. According to second order perturbation
theory within the Schr\"odinger picture, the expression for the second
order wavefunction is 
\begin{equation}
\begin{split}\left|\psi^{\left(2\right)}\left(t\right)\right\rangle =-\frac{1}{4\hbar^{2}} & \int_{-\infty}^{t}dt_{2}\int_{-\infty}^{t_{2}}dt_{1}\\
 & e^{-\frac{i}{\hbar}\widehat{H}_{0}\left(t-t_{2}\right)}\left(\widehat{\mu}E_{S}e^{i\omega_{S}t_{2}}\right)\times\\
 & e^{-\frac{i}{\hbar}\widehat{H}_{0}\left(t_{2}-t_{1}\right)}\left(\widehat{\mu}E_{I}e^{-i\omega_{I}t_{1}}\right)\times\\
 & e^{-\frac{i}{\hbar}\widehat{H}_{0}t_{1}}\left|\psi_{i}\right\rangle ,
\end{split}
\label{eq:2nd_order_wf}
\end{equation}
where the initial state of the system is $\left|\psi_{i}\right\rangle $.
Here $\widehat{H}_{0}$ is the unperturbed Hamiltonian of the electronic
system and $\widehat{\mu}$ is the transition dipole operator. \footnote{Formally, there is also another contribution to Eq.~\eqref{eq:2nd_order_wf}
with the order of the operators $\hat{\mu}E_{S}e^{i\omega_{S}t}/2$
and $\hat{\mu}E_{I}e^{-i\omega_{I}t}/2$ switched. This term leads
to another contribution to $\alpha_{fi}$ in Eq.~\eqref{eq:alpha_fi}
which appears in the standard frequency-domain KHD formula.\citep{albrecht_theory_1961,heller_simple_1982}
However, this additional term corresponds to a relatively unlikely
process whereby the system first emits an outgoing photon, then absorbs
an incoming photon. For resonant Raman scattering, the contribution
of this term can be ignored.}

Now we would like to express the number of outgoing photons scattered
per unit time in terms of the change in the second order wavefunction.
To do so, we evaluate the time derivative of the second-order wavefunction
and insert a complete set of final states $\left|\psi_{f}\right\rangle $
to obtain:
\begin{equation}
\frac{\mathrm{d}}{\mathrm{d}t}\left\Vert \psi^{\left(2\right)}\left(t\right)\right\Vert ^{2}=\frac{2\pi E_{S}^{2}E_{I}^{2}}{16\hbar^{2}}\sum_{f}\left|\alpha_{fi}\left(\omega_{I}\right)\right|^{2}\delta\left(\omega_{S}-\Delta\omega\right)\label{eq:2nd_order_wf_0}
\end{equation}
where the frequency-dependent polarizability is defined by 
\begin{equation}
\alpha_{fi}\left(\omega_{I}\right)=\frac{i}{\hbar}\int_{0}^{\infty}d\tau\left\langle \psi_{f}\right|\widehat{\mu}e^{-\frac{i}{\hbar}\widehat{H}_{0}\tau}\widehat{\mu}e^{i\left(\omega_{I}+\omega_{i}\right)\tau}\left|\psi_{i}\right\rangle .\label{eq:alpha_fi}
\end{equation}
Here $\Delta\omega=\omega_{I}+\omega_{i}-\omega_{f}$, and $\hbar\omega_{i}$
and $\hbar\omega_{f}$ are the energy levels of the initial and final
states of the system. If we now invoke the 1D density of states for
photons ($\rho\left(\omega_{S}\right)=\frac{L}{\pi c}$), we can eliminate
the delta function in Eq.~\eqref{eq:2nd_order_wf_0} and write
\begin{equation}
\frac{\mathrm{d}}{\mathrm{d}t}\left\Vert \psi^{\left(2\right)}\left(t\right)\right\Vert ^{2}=\frac{L}{8\hbar^{2}c}E_{S}^{2}E_{I}^{2}\left|\alpha_{fi}\left(\omega_{I}\right)\right|^{2}.\label{eq:2nd_order_wf_1}
\end{equation}

Lastly, in order to express the scattering cross section in terms
of photon frequencies, we must calculate the amplitude of the scattered
EM field in Eq.~\eqref{eq:2nd_order_wf_1} in terms of other physical
observables. To do so ,we note the simple and general relationship
between the electric field amplitude $E$ and the number of photon
$N$ in a volume $L$:
\begin{equation}
\frac{E^{2}}{2}=\hbar\omega\frac{N}{L}.\label{eq:Einstein-photon}
\end{equation}
Here $\frac{N}{L}$ is the photon density for a 1D system. Note that
Eq.~\eqref{eq:Einstein-photon} is valid for both incoming and scattered
photons. For incoming photons, the incident field intensity satisfies
\begin{equation}
E_{I}^{2}=2\hbar\omega_{I}\frac{N_{I}}{L}.
\end{equation}
For scattered photons, assuming spontaneous emission, we must have
$N_{S}=1$ such that 
\begin{equation}
E_{S}^{2}=\frac{2\hbar\omega_{S}}{L}.
\end{equation}
With these relations, we rewrite Eq.~\eqref{eq:2nd_order_wf_1} as:
\begin{equation}
\frac{\mathrm{d}}{\mathrm{d}t}\left\Vert \psi^{\left(2\right)}\left(t\right)\right\Vert ^{2}=\frac{\omega_{I}\omega_{S}}{2c}\frac{N_{I}}{L}\left|\alpha_{fi}\left(\omega_{I}\right)\right|^{2}.\label{eq:2nd_order_wf_2}
\end{equation}

Finally, we divide Eq.~\eqref{eq:2nd_order_wf_2} by the incident
photon flux ($\frac{N_{I}c}{L}$) and obtain the Raman scattering
cross section for a 1D system:

\begin{equation}
\sigma_{fi}^{\mathrm{1D}}\left(\omega_{S},\omega_{I}\right)=\frac{\omega_{I}\omega_{S}}{2c^{2}}\left|\alpha_{fi}\left(\omega_{I}\right)\right|^{2}.
\end{equation}

\section{Coherent emission intensity of Raman scattering\label{sec:Coherent-emission-intensity}}

Here, we derive the coherent emission intensity of a three-level system
within the rotating wave approximation (RWA). We let $\left|1_{\omega}\right\rangle $
be a state of the EM field with one photon of mode $\omega$, and
denote the vacuum state as $\left|\left\{ 0\right\} \right\rangle $.
The dressed state representation of the total wavefunction can be
written as 
\begin{equation}
\left|\psi\left(t\right)\right\rangle =\sum_{j=0,1,2}C_{j,0}\left(t\right)\left|j;\left\{ 0\right\} \right\rangle +\sum_{j=0,1,2}C_{j,\omega}\left(t\right)\left|j;\omega\right\rangle .
\end{equation}
Here the basis consists of $\left|j;\left\{ 0\right\} \right\rangle =\left|j\right\rangle \left|\left\{ 0\right\} \right\rangle $
and $\left|j;\omega\right\rangle =\left|j\right\rangle \left|1_{\omega}\right\rangle $
including up to a single photon per mode. For the incoming photon
of mode $\omega_{I}$, we choose the initial state is to be
\begin{equation}
\left|\psi\left(0\right)\right\rangle =C_{0,0}\left|0;\left\{ 0\right\} \right\rangle +C_{0,\omega_{I}}\left|1;\omega_{I}\right\rangle .
\end{equation}
with $\left|C_{0,0}\right|^{2}+\left|C_{0,\omega_{I}}\right|^{2}=0$.
Here, we are approximating a weak coherent state as the sum of zero
and one photon states only.

Now we assume that the incoming field is at resonance with states
$\left|0\right\rangle $ and $\left|2\right\rangle $, i.e. $\hbar\omega_{I}=\varepsilon_{2}-\varepsilon_{0}$.
The Raman scattering frequency is $\hbar\omega_{R}=\varepsilon_{2}-\varepsilon_{1}$
and the Rayleigh scattering frequency is $\hbar\omega_{I}$. Within
the RWA, we consider the resonant states $\left|0;\omega_{I}\right\rangle $,
$\left|1;\omega_{R}\right\rangle $, and $\left|2;\left\{ 0\right\} \right\rangle $
and write the RWA Hamiltonian as 
\begin{equation}
{\cal H}_{\mathrm{RWA}}=\left(\begin{array}{ccc}
\varepsilon_{0}+\hbar\omega_{I} & 0 & V_{02}\\
0 & \varepsilon_{1}+\hbar\omega_{R} & V_{12}\\
V_{02}^{*} & V_{12}^{*} & \varepsilon_{2}
\end{array}\right)
\end{equation}
where $V_{02}=-\mu_{02}E$ and $V_{12}=-\mu_{12}E$. In addition to
the resonant states, the excited state $\left|2\right\rangle $ is
coupled to the continuous manifolds, $\left\{ \left|0;\omega\right\rangle ,\left|1;\omega\right\rangle \right\} $.
Therefore, the steady state solution can be expressed in terms of
the resonant states and the initial vacuum state:
\begin{equation}
\begin{split}\left|\overline{\psi_{\mathrm{RWA}}}\right\rangle = & \overline{C_{0,0}}\left|0;\left\{ 0\right\} \right\rangle +\overline{C_{0,\omega_{I}}}\left|0;\omega_{I}\right\rangle +\\
 & \overline{C_{1,\omega_{R}}}\left|1;\omega_{R}\right\rangle +\overline{C_{2,0}}\left|2;\left\{ 0\right\} \right\rangle +\\
 & \sum_{\omega\neq\omega_{I}}\overline{C_{0,\omega}}\left|0;\omega\right\rangle +\sum_{\omega\neq\omega_{I}}\overline{C_{1,\omega}}\left|1;\omega\right\rangle .
\end{split}
\end{equation}
Note that the vacuum state $\left|0;\left\{ 0\right\} \right\rangle $
is not coupled to the resonant states. 

Finally, we can evaluate the coherent emission intensity using the
expectation value of the electric field. For a 1D system, the electric
field operator is given by 
\begin{equation}
\widehat{E}\left(x\right)=i\sum_{k}{\cal E}_{k}\left(\widehat{a}_{k}e^{ikx}-\widehat{a}_{k}^{\dagger}e^{-ikx}\right)
\end{equation}
with ${\cal E}_{k}=\sqrt{\frac{\hbar\omega_{k}}{2\epsilon_{0}L}}$
in a space of volume $L$. Using the form of the steady state wavefunction,
we can obtain the lowest order approximation of the expectation value
of the electric field:
\begin{equation}
\left\langle \widehat{E}\left(x\right)\right\rangle =-2{\cal E}_{k}\mathrm{Im}\left(\overline{C_{0,0}}^{*}\overline{C_{0,\omega_{I}}}e^{i\omega_{I}x/c}\right).\label{eq:RWA_Efield}
\end{equation}
We note that there is not any contribution to the electric field at
frequency $\omega_{R}$ since $\left|\left\langle 1;\left\{ 0\right\} \right|\overline{\psi_{\mathrm{RWA}}}\right\rangle =0$.
Thus, the Fourier transform of the electric field vanishes at the
Raman frequency,
\begin{equation}
\left|\left\langle E\left(\omega_{R}\right)\right\rangle \right|=0.
\end{equation}
In other words, within the RWA, resonant Raman scattering does not
yield coherent emission. 

Therefore, we must conclude that, for a more general situation not
far from RWA, all Raman scattering signals must be dominate by incoherent
emission. As a sidenote, the arguments above also show that the Rayleigh
peak should be coherent: the electric field in Eq.~\eqref{eq:RWA_Efield}
does not vanish at frequency $\omega_{S}=\omega_{I}$.\bibliographystyle{apsrev4-1}
\bibliography{MyLibrary}

\end{document}